\documentclass[10pt,letterpaper]{article}
\usepackage{opex3}

\usepackage{color}
\usepackage{amsmath}

\newcommand{\NYVO}{Nd:YVO$_4$}

\usepackage{upgreek} 

\usepackage[backref=page]{hyperref}
\hypersetup{
		colorlinks=true,	      
	  linkcolor=blue,          
    citecolor=blue,        
    urlcolor=green,           
}

\begin{document}

\title{2.1-watts intracavity-frequency-doubled all-solid-state light source at 671\,nm for laser cooling of lithium}

\author{Ulrich Eismann$^1$, Andrea Bergschneider$^{1,2}$, Christophe Salomon$^1$ and Fr\'ed\'eric Chevy$^1$}

\address{$^1$Laboratoire Kastler Brossel, ENS, UPMC, CNRS UMR 8552 \\ 24 rue Lhomond, 75231 Paris, France\\
$^2$Physikalisches Institut, Ruprecht-Karls-Universit\"at Heidelberg \\ Im Neuenheimer Feld 226, 69120 Heidelberg, Germany}

\email{ueismann@gmail.com} 



\begin{abstract}
We present an all-solid-state laser source emitting up to 2.1\,W of single-frequency light at 671\,nm developed for laser cooling of lithium atoms. It is based on a diode-pumped, neodymium-doped orthovanadate (Nd:YVO$_4$) ring laser operating at 1342\,nm. Optimization of the thermal management in the gain medium results in a maximum multi-frequency output power of 2.5\,W at the fundamental wavelength. We develop a simple theory for the efficient implementation of intracavity second harmonic generation, and its application to our system allows us to obtain nonlinear conversion efficiencies of up to 88\%. Single-mode operation and tuning is established by adding an etalon to the resonator. The second-harmonic wavelength can be tuned over 0.5\,nm, and mode-hop-free scanning over more than 6\,GHz is demonstrated, corresponding to around ten times the laser cavity free spectral range. The output frequency can be locked with respect to the lithium $D$-line transitions for atomic physics applications. Furthermore, we observe parametric Kerr-lens mode-locking when detuning the phase-matching temperature sufficiently far from the optimum value.
\end{abstract}


\bibliographystyle{osajnl}

\section{Introduction}
\label{s:intro}

Lithium atoms are one of the most versatile species used for research on quantum gases. Nature offers significantly abundant bosonic and fermionic isotopes to the experimentalist, allowing the study of both types of quantum statistics\,\cite{Giorgini2008} in the atomic physics lab. The simple yet powerful technique of magnetic Feshbach tuning of the atomic interactions\,\cite{Chin2010} is the quintessential ingredient for a large number of experiments, and it explains the outstanding role the lithium atom has played in the development of the field. 

For realizing a degenerate quantum gas, one needs to implement different laser cooling schemes which typically require near-resonant single-frequency light input in the watt range. The benefits of all-solid-state designs like large output power, high reliability, low maintenance effort and high intrinsic stability are helpful for realizing light sources that will be welcome tools for every ultracold atom experiment.

We will present in this article an all-solid-state laser emitting multi-watt single-frequency radiation near \mbox{671\,nm}. The diode-pumped design is based on neodymium-doped orthovanadate (\NYVO) as the gain medium, lasing at the  fundamental wavelength of 1342\,nm. Second harmonic generation (SHG) is then established using periodically-poled potassium titanyl phosphate (ppKTP) as the nonlinear medium. The development of a number of related laser sources has been published recently\,\cite{Hou2011,Camargo2010,Eismann2012}, and the system developed in our group\,\cite{Eismann2012} has proven highly reliable in every-day operation over the period of one year. However, these sources are currently limited to output powers in the few-hundred-milliwatts regime. We have identified two major limitations for pushing the \NYVO-SHG concept into the multi-watt range, which are the loss introduced by thermal lensing in the gain medium, and the nonlinear conversion efficiency. Therefore, the key ingredients of our novel high-power design are an improved heat management in the \NYVO, and the implementation of intracavity SHG, both of which will be discussed extensively. Furthermore, we carefully specify the important parameters of the laser output and find that it largely satisfies the exigencies of laser cooling. Our source has successfully been used in the implementation of a 
gray-molasses cooling scheme for \mbox{lithium} atoms, see Ref.\,\cite{Fernandes2012}. 

Apart from laser cooling of atoms, more applications in the fields of atomic physics and nonlinear optics are currently limited by the available single-frequency 671-nm power. In atom interferometers, a larger and thus more homogeneous gaussian beam can increase the signal-to-noise ratio. The large spatial splitting of the atomic wavepackets employing the lightweight lithium species is favorable\,\cite{Miffre2006}, and can even be increased using multi-photon Bragg scattering\,\cite{Muller2008} when high-intensity laser beams are employed. Furthermore, the lithium $D$-line isotope splitting is large enough to allow selective addressing of the isotopes in hot vapors, making a narrow-bandwidth source attractive for lithium isotope separation\,\cite{Olivares2002}. In addition, the creation and long-distance transmission of entangled photons in the low-dispersion, low-absorption wavelength region of standard silica fibers near 1.3\,$\upmu$m has recently been proposed. The scheme uses the output of a sub-threshold optical parametric oscillator in the degenerate regime pumped by a single-frequency 671-nm laser\,\cite{Hou2011}. Finally, our laser could serve as a low-intensity-noise pump for Cr:LiSAF lasers\,\cite{Payne1994}.

The article is organized as follows: In Section\,2 we present the optimization of the fundamental laser design. In Section\,3, we implement intracavity second harmonic generation. In Section\,4, we report on the course and continuous fine tuning behavior and nonlinear-Kerr-lens mode locking, and we conclude in Section\,5.

\section{The fundamental laser}  
\label{s:IRSetup}

A first step towards a stable high-power frequency doubled laser source is the availability of an efficient laser system at the fundamental wavelength. To minimize detrimental thermal effects in the gain medium, the following pathway has been chosen: A pump wavelength of $888\,\rm nm$ in contrast to the former $808\,\rm nm$\,\cite{Eismann2012} leads to a lower quantum defect per absorption-emission cycle\,\cite{McDonagh2006}, and thus to lower heating for a given pump rate. In addition, a larger value for the laser crystal length has been chosen in order to spread the heat input over a bigger volume. Therefore, heat transport from the central region to the crystal mount is facilitated. The \NYVO  peak temperature is lower, and thermal issues are less of a concern.

\begin{figure}[htbp]
\centering\includegraphics[width=\linewidth]{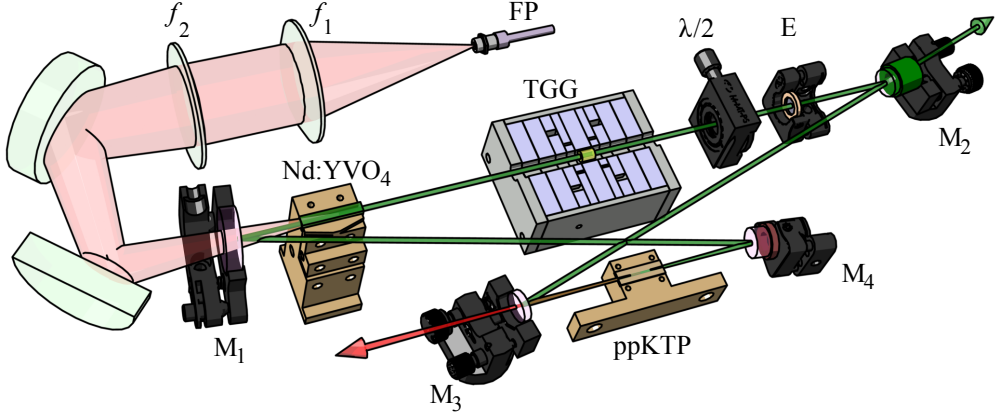}
\caption{(Color online) The laser setup. The pump source, a fiber-coupled diode laser bar\,(FP), is imaged into the gain medium by a pair of lenses $f_1$ and $f_2$. The Nd:YVO$_4$ gain medium is placed in a four-mirror bow-tie ring resonator consisting of mirrors M$_{1-4}$, which are highly reflecting at 1342\,nm. Unidirectional operation is forced employing a terbium gallium garnet (TGG)-based Faraday rotator in combination with a half-wave plate ($\lambda/2$). The use of an etalon (E) allows for stable single-longitudinal-mode operation. The nonlinear crystal (ppKTP) is inserted at the tight focus between the curved mirrors M$_3$ and M$_4$. The second harmonic output beam (red) is transmitted through M$_3$. For the measurements presented in Section\,\ref{s:IRSetup}, the ppKTP was removed and the distance M$_3$-M$_4$ adjusted accordingly, and the high-reflectivity mirror M$_2$ was replaced by a partly transmitting output coupling mirror. The fundamental laser beam (green) is then coupled out through M$_2$.}
\label{f:laser}
\end{figure}

A schematic overview of the laser setup is given in Fig.\,\ref{f:laser}. The output of an 888-nm fiber-coupled diode laser bar (NA = 0.22, $400\,\upmu$m fiber core diameter) is imaged by two lenses ($f_1 = 75\,$mm, $f_2 = 200\,$mm) into the 1.0\,\%\,at.-doped \NYVO~crystal. The crystal (a-cut, $4\times4\times25\,{\rm mm}^3$, anti-reflection coated on both sides for $1342\,\rm nm$ and $888\,\rm nm$) is wrapped in indium foil and mounted in a water-cooled copper block. The mirrors M$_{1-4}$ constitute a bow-tie cavity. M$_1$, M$_3$ and M$_4$ are highly reflective at $1342\,\rm nm$, and M$_1$ is transmitting at 888\,nm. M$_3$ and M$_4$ are concave mirrors with a radius of curvature of $100\,\rm mm$. 
M$_2$ is the output coupler for which mirrors with different values of transmission are available. The cavity dimensions are M$_1$M$_2 \approx 300\,\rm mm$, M$_2$M$_3 \approx $M$_1$M$_4 \approx 210\,\rm mm$ and M$_3$M$_4 \approx 97\,\rm mm$. For forcing unidirectional operation we use a Faraday rotator consisting of a terbium-gallium-garnet rod-shaped crystal (TGG) of $6\,\rm mm$ length embedded in a strong permanent magnet \cite{Tr'enec2011} in combination with a true-zero-order half-wave plate. An uncoated infrared fused silica etalon of $500\,\rm \upmu m$ thickness 
serves as a wavelength selective element.

\begin{figure}[htbp]
\centering\includegraphics[width=\linewidth]{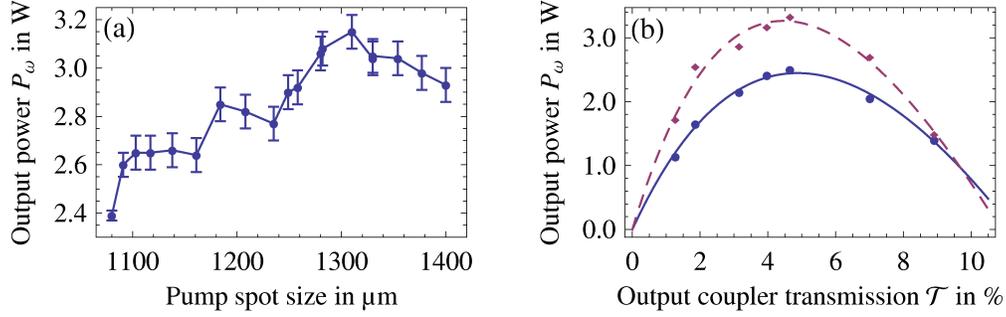}
\caption{(Color online) (a)~Optimization of the output power by changing the pump spot diameter, performed on the laser cavity presented in Fig.\,\ref{f:laser} with all the intracavity elements removed, except for the \NYVO. For a $\mathcal{T} = 5\%$ output coupler (M$_2$), the mode overlap was optimized for each pump spot diameter by slight adjustments of the curved-mirror distance M$_3$-M$_4$. Lines are guides to the eye only. (b)~Rigrod analysis. The infrared output power $P_\omega$ is measured as a function of the output coupler transmission $\mathcal{T}$ and fitted with the Rigrod model~\eqref{e:rigrod} for bidirectional (red diamonds, dashed line) and unidirectional (blue circles, solid line) operation at $P_{\rm abs,max} = 32.5\,\rm W$ and optimized mode overlap. In both cases the optimum transmission is found at $\mathcal{T} \approx 5\%$. The parasitic roundtrip loss determined from the fits yield $10(4)\%$ for the bidirectional and $\mathcal{L}= 16(6)\%$ for the unidirectional case.}
\label{f:OverlapAndRigrod}
\end{figure}

For maximum power output, it is crucial to optimize the overlap of the pump beam and the cavity mode  \cite{Laporta1991, Chen1997}. For simplicity, we perform this operation on the empty laser cavity, consisting of M$_{1-4}$ and the \NYVO  only. We apply the maximum value of the absorbed pump power $P_{\rm abs,max}=32.5\,\rm W$, and a $\mathcal{T}=5\%$-transmission output coupler (M$_{2}$). By changing the magnification of the pump imaging setup consisting of $f_1$ and $f_2$, the top-hat shaped pump spot diameter was altered between $1080\,\rm{\upmu m}$ and $1400\,\rm{\upmu m}$, see Fig.\,\ref{f:OverlapAndRigrod}. The size of the cavity mode in the laser crystal can be changed using the curved mirror distance M$_3$M$_4$, and was optimized for each data point. The maximum output power is obtained at a pump diameter of around $1300\,\rm{\upmu m}$, where it is kept for the remainder of this article.

For both the bidirectional (empty cavity only containing \NYVO) and the unidirectional (additional TGG and half-wave plate) operation we measured the maximum output power as a function of the output coupler transmission, see Fig.\,\ref{f:OverlapAndRigrod}. 
In both cases a $\mathcal{T} = 5\,\%$ mirror delivers the maximum fundamental output power $P_{\omega}$. By equating the single-pass gain with the total round-trip loss $\mathcal{L}_{\rm tot} = \mathcal{L} + \mathcal{L}_{\rm out} = \mathcal{L} + \mathcal{T}$ for both the bi- and unidirectional cases, where $\mathcal{L}$ is the sum of the parasitic round-trip losses, we find
\begin{eqnarray} \label{e:rigrod}
P_{\omega} = P_{\rm sat} \mathcal{T} \left[\frac{G_0}{\mathcal{T}+\mathcal{L}} -1\right]\,\rm 
\end{eqnarray}
similarly to~\cite{Rigrod1963}, where  $P_{\rm sat}$ is the saturation power, $G_0$ the small-signal gain and $\mathcal{L}$ the sum of the parasitic round-trip losses. Although this analysis essentially relies on plane waves, it can be mapped to the more realistic case of top-hat pump beams and gaussian laser cavity eigenmodes, as present in diode-pumped solid-state lasers, see
~\cite{Eismann2012} and references therein.
A least-squares fit to~\eqref{e:rigrod} yields $\mathcal{L} = 10(4)\%$ for the parasitic loss in the bidirectional case. In~\cite{Eismann2012} we found the loss in an empty four-mirror bow-tie cavity of mirrors from the same batch to be smaller than 1\%. We attribute the remaining $9(4)\%$ mainly to aberrations caused by the thermal lens in the \NYVO. 

In the unidirectional case, the fitting procedure yields $\mathcal{L} = 16(6)\%$. Compared to the bidirectional case, the difference of $\mathcal{L}_{\rm Faraday}=6(7)\%$ can be attributed to the insertion of the TGG crystal and the waveplate. Indeed, thermal depolarization and the accompanying loss is a well-known effect in TGG. It imposes stringent limits on power scaling of unidirectional ring lasers and Faraday isolators\,\cite{Khazanov2004}. We observe a dependence of the optimum half-wave plate angle on the circulating intracavity power, which is strong evidence of this effect. From the fit, we furthermore obtain the values $P_{\rm sat} = 170(50)$\,W and $G_0 = 0.27(5)$ for unidirectional operation, which are important for the optimization of intracavity doubling, see Section\,\ref{s:ICSHG}. 

\begin{figure}[htbp]
\centering\includegraphics[width=7cm]{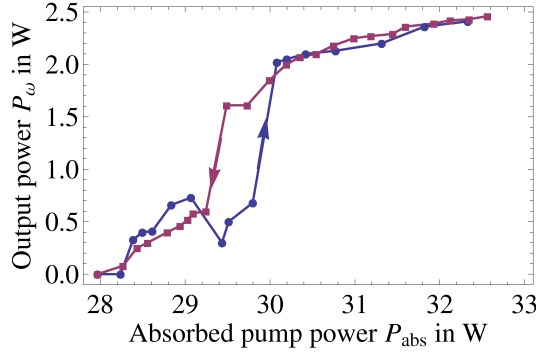}
\caption{(Color online) Infrared unidirectional output power $P_\omega$ as a function of the absorbed pump power $P_{\rm abs}$. The setup is optimized for the maximal absorbed pump power $P_{\rm abs,max} = 32.5$\,W. The oscillation threshold is found at $P_{\rm abs}\approx 28\,\rm W$. The data shows hysteresis between $P_{\rm abs} = 29\,\rm W$ and 30\,W, as indicated by the arrows for increasing or decreasing pump power. This behavior is typical for high-power designs. After a sudden rise the output power increases only slowly until it eventually reaches the maximum of $2.5\,\rm W$ at $P_{\rm abs,max}$. Lines are guides to the eye only.}
\label{f:PPerformanceIR}
\end{figure}

For the now-optimized unidirectional infrared setup, we measure the output power as a function of the absorbed pump power $P_{\rm abs}$, see Figure\,\ref{f:PPerformanceIR}. As the optimization is performed at the maximum absorbed pump power  $P_{\rm abs, max}\approx 32.5$\,W, the laser emission only starts at $P_{\rm abs} \approx 28$\,W. After reaching threshold, the output power is unstable and displays a hysteresis feature between $P_{\rm abs} = 29$\,W and 30\,W. After crossing the hysteresis region, the laser emission is stable and only weakly depends on $P_{\rm abs}$. This behavior is typical for high-power solid-state laser designs and has been reported in~\cite{Lenhardt2009}. 
As discussed before, thermal depolarization in the TGG is significant and can lead to a change of the lasing direction and occasional bistable behavior when increasing the pump power. Thus, the angle of the half-wave plate had to be adjusted for every data point presented here. We keep the pump power constant at $P_{\rm abs, max}\approx 32.5$\,W in the remainder of the article.

\section{Efficient intracavity second-harmonic generation}
\label{s:ICSHG}
Efficient frequency doubling  of infrared lasers can be established using periodically-poled nonlinear crystals in an external cavity. Using this method at a fundamental wavelength of 1342\,nm, a doubling efficiency of $P_{2\omega}/P_{\omega} = 86\%$ has been obtained in our first-generation setup\,\cite{Eismann2012}, and serves as a benchmark. However, a more direct approach followed here is intracavity second harmonic generation (ICSHG), which requires only one cavity and thus represents an important simplification.

For the analysis of the output power of ICSHG lasers, the ouput coupling loss $\mathcal{L}_{\rm out} = \mathcal{T}$ discussed in Section\,\ref{s:IRSetup} needs to be replaced by $\eta P$, where $\eta$ is the single-pass doubling efficiency, and $P$ the circulating intracavity power. Similar to the one found by R.G.~Smith\,\cite{Smith1970}, the solution for the SH output power in the unidirectional case reads
\begin{eqnarray} \label{e:smith}
P_{2\omega} = \frac{P_{\rm sat} G_0}{\xi} \left[
 \sqrt{(\xi-\zeta)^2 +\xi} -(\xi + \zeta) \right]^2\,\rm ,
\end{eqnarray}
where $\xi=\eta P_{\rm sat}(4G_0)^{-1}$ and $\zeta=\mathcal{L}(4G_0)^{-1}$ are the dimensionless output coupling and loss parameters. As pointed out in~\cite{Smith1970}, it is interesting to note that the value for the optimum output coupling $\mathcal{L}_{\rm out, opt} = \sqrt{G_0\mathcal{L}} - \mathcal{L}$ is the same for both linear and non-linear output coupling mechanisms, and delivers the same amount of output power. However, the round-trip parasitic loss $\mathcal{L}$ will contain an additional contribution from the insertion of the nonlinear medium, such that $P_{2\omega}/P_{\omega} < 1$ for any realistic system. Using the fit values from Section\,\ref{s:IRSetup}, we maximize \eqref{e:smith} by choosing an optimum single-pass doubling efficiency $\eta_{\rm opt} = 0.10(5)\%.$W$^{-1}$.

To evaluate $\eta$, we refer to the Boyd-Kleinman theory for focused gaussian beams\,\cite{Boyd1968},
\begin{equation}
\eta(T) = \frac{2\omega^3 d_{ij}^2L}{\pi\varepsilon_0 c^4 n_{\omega,i}(T) n_{2\omega,j}(T)}\times h[\alpha,\beta(T)]\,\mathrm{,}
\label{e:BK}
\end{equation}
where $\varepsilon_0$ is the vacuum permittivity, $c$ the speed of light in vacuum, $d_{ij}$ is the $i,j$-th element of the material's nonlinear tensor, $n_{\omega,i}(T)$ the material's refractive index along the $i$~axis at angular frequency $\omega$ and temperature $T$, and $L$ the nonlinear material's length. $d_{ij}$ needs to be replaced by $d_{\rm eff} = 2 d_{ii}\pi^{-1}$ for periodically poled (pp) materials. The function
\begin{equation}
h[\alpha,\alpha_0,\beta(T)] = \frac{1}{4\alpha} \left| \int\limits_{-\alpha-\alpha_0}^{\alpha-\alpha_0} \! \frac{{\rm e}^{\mathrm{i}\beta(T)\tau}}{1 + \mathrm{i}\tau} \, \mathrm{d}\tau \right|^2\,\mathrm{,}\label{e:hfunc}
\end{equation}
is the dimensionless Boyd-Kleinman function\,\cite{Boyd1968}. $\alpha = L\,(2z_{\rm R})^{-1}$ and $\alpha_0 = z_0\,(2z_{\rm R})^{-1}$ are the focusing and offset parameters, respectively. $z_{\rm R}$ is the Rayleigh length, and $z_0$ is the offset of the beam focus with respect to the nonlinear medium's center.
%
%
%
%
The phase-matching parameter reads $\beta(T) = 4 \pi z_{\rm R}\lambda^{-1}\times \left\lbrace  n_{\omega,i}(T) - n_{2\omega,j}(T) - \lambda[2\Lambda(T)]^{-1} \right\rbrace $, where $\lambda$ is the vacuum wavelength. The $\Lambda(T)$ term only occurs for periodically poled materials, where $\Lambda(T)$ is the poling period, and the temperature dependence results from thermal expansion.

In the intracavity-doubling setup, we replace the output coupling mirror M$_2$ by a high-reflectivity mirror, cf. Figure\,\ref{f:laser}. The nonlinear, periodically-poled potassium titanyl phosphate crystal (ppKTP) is inserted between the concave mirrors M$_3$ and M$_4$. To account for the {ppKTP} refractive index, the distance M$_3$M$_4$ is increased to $\approx $106\,\rm mm. The nonlinear crystal has outer dimensions of $1 \times 6 \times 18\,\rm mm^3$ and is antireflection coated at 1342\,nm and 671\,nm. Its poling period $\Lambda = 17.61\,\upmu$m is chosen for phase matching at 23.5$^\circ$C according to the temperature-dependent Sellmeier equations from\,\cite{Fradkin1999,Emanueli2003}. However, in a previous study we found the phase-matching temperature at 33.2$^\circ$C\,\cite{Eismann2012}. This deviation can be explained by uncertainties on the reported Sellmeier equations, and the manufacturing tolerance on the ppKTP poling period. The nonlinear coefficient $d_{33} = 14.5$pm.V$^{-1}$ found for the crystal in our preliminary study is 14\% lower than the highest literature value reported so far\,\cite{Peltz2001}, probably due to poling imperfections\,\cite{Eismann2012}. Using this crystal for nonlinear output coupling, we can easily obtain a single-pass efficiency $\eta > \eta_{\rm opt}$, such that design considerations are relaxed, and we are able to tune to $\eta_{\rm opt}$ by changing the phase-matching temperature. For temperature stabilization, the crystal is wrapped in indium foil and mounted in a temperature-controlled copper block. The frequency-doubled light is transmitted through mirror M$_3$.

\begin{figure}[]
\centering\includegraphics[width=8cm]{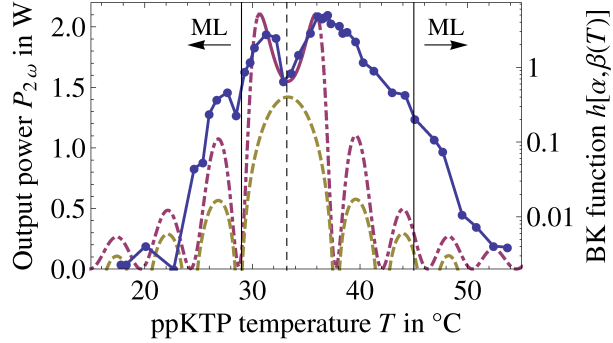}
\caption{(Color online) Output power as a function of the phase-matching temperature (points, blue lines are a guide to the eye only). The data shows a double-peak structure of 1.9\,W / 2.1\,W of output power slightly off of the optimum phase-matching temperature of 33.2$^\circ$C (central vertical line). A simple theoretical model presented in the text (dash-dotted purple line) describes the data well in the central high-conversion region, using the known temperature dependence of the single-pass doubling efficiency, which is proportional to the dimensionless Boyd-Kleinman function $h(T)$ [dashed gold line, Eq.\,\eqref{e:hfunc}]. The dashed vertical line indicates the perfect phase-matching temperature of $T_{\rm pm} = 33.2^\circ$C, where $\beta(T_{\rm pm}) \simeq 0$. The vertical lines with arrows indicate the temperature regions where self-mode locking occurs, cf. Section\,\ref{s:FB}.}
\label{f:ppKTP-T-tuning}
\end{figure}

In Fig.\,\ref{f:ppKTP-T-tuning} we present the measurement of the second-harmonic output power as a function of the phase-matching temperature of the ppKTP crystal. We find a maximum of 2.1\,W of output power at $T_{\rm pm} = 36^\circ$C, and a second maximum of 1.9\,W at $T_{\rm pm} = 31^\circ$C. As expected from Eq.\,\eqref{e:smith}, at the perfect phase matching temperature $\eta > \eta_{\rm opt}$, and the output power only amounts to 1.6\,W. The FWHM of the temperature-tuning curve amounts to 22\,K. As compared to the maximum infrared power presented in the former section, we obtain a maximum doubling efficiency of 88\%. The excellent mode quality of the second-harmonic light is confirmed by a single-mode fiber coupling efficiency larger than 80\%. 

To gain deeper insight in our data, we employ the ICSHG theory (Eqs.~\eqref{e:smith}-\eqref{e:hfunc}). We include the additional parasitic intracavity loss due to the presence of the etalon and the nonlinear crystal, which amounts to $\mathcal{L}_{\rm add} = 0.6(1)\%$. Thus, the theory predicts a maximum SH output power of 2.1\,W, as is found experimentally. The value of $\mathcal{L}_{\rm add}$ is compatible to the sum of the ppKTP insertion loss measured independently\,\cite{Eismann2012} and the calculated walk-off loss for the etalon\,\cite{Eismann2012a}. From an ABCD-matrix formalism we obtain the cavity eigenmode, yielding $z_{\rm R} = 11.5\,mm$ and $z_0 = 3\,mm$. 
At perfect phase matching, we obtain $h = 0.40$, yielding a SH output power of 1.6\,W, which is in excellent agreement with our findings. This gives us further confidence in our previously measured value of $d_{33}$\,\cite{Eismann2012}.
The temperature-dependent KTP refractive indices and the thermal expansion coefficient of the poling period were presented in~\cite{Fradkin1999,Emanueli2003}. We adjust $\beta$ by an additive constant in order to account for the measured phase matching condition $\beta(T = 33.2^\circ{\rm C}) = 0$, as justified before.  The theory (solid purple line in Fig.\,\ref{f:ppKTP-T-tuning}) describes the data well in the central high conversion region, and yields the characteristic double-peak structure. 

Outside of the central region, the circulating intracavity power rises significantly and the power-dependent thermal depolarization loss would need to be accounted for in our simple model. As compared to the theory, the data displays less evidence of dips. This can be explained by a thermal smearing effect due to residual absorption in the ppKTP at large powers, yielding a spatial dependence of the phase-matching parameter $\beta(T)$. Outside of the region indicated by the solid vertical lines, the laser is in mode-locked (pulsed) operation, and we do not expect our model to be applicable.

\section{Tuning behavior and nonlinear-Kerr-lens mode locking}\label{s:FB}

For course tuning and single-longitudinal-mode (SLM) operation, the laser was equipped with an uncoated etalon of 0.5\,mm length, yielding a free spectral range of 210\,GHz. We note that this single, weakly-selective etalon is sufficient for this purpose due to the self suppression of axial mode hopping in intracavity-frequency-doubled lasers\,\cite{Martin1997,Helmfrid1994}. 

\begin{figure}[htbp]
\centering\includegraphics[width=10cm]{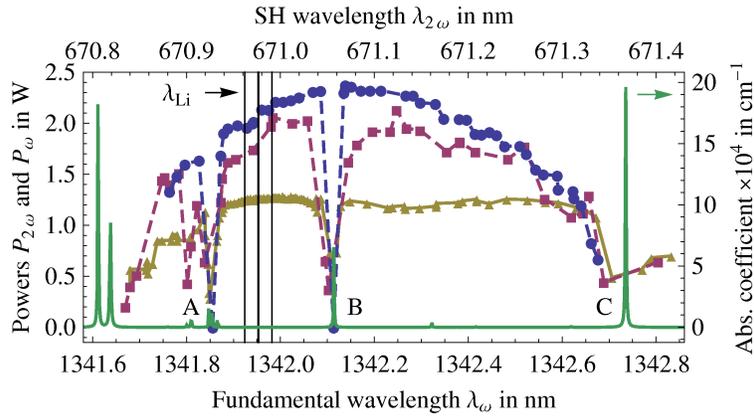}
\caption{(Color online, lines are guides for the eye only) Single-frequency output spectra of the infrared laser presented in Section\,\ref{s:IRSetup} (blue circles), the intracavity-frequency-doubled laser (purple squares) and the infrared source presented in\,\cite{Eismann2012} (gold triangles). For easy comparison, all wavelengths are given in vacuum values. 
The vertical lines denote the positions of the lithium-$D$ line resonances. The green line shows a water vapor absorption spectrum for typical parameters (23\,$^\circ$C, 60\% rel. humidity). The wavelength regions marked A,B,C where stable, powerful operation of the lasers can not be established coincide with absorption peaks of water molecules. The level of output power of the infrared laser and the frequency-doubled laser are closely spaced, proving that the nonlinear crystal introduces weak additional passive loss in the laser cavity, whereas the degree of nonlinear output coupling is at its optimum value.}
\label{f:tuningcurve}
\end{figure}

The tuning range of the frequency-doubled laser is characterized and compared to the infrared laser in Fig.\,\ref{f:tuningcurve}. 
Apart from having the same shape and features, it is striking to note that the output power of the frequency-doubled laser amounts to almost the same value as the non-doubled laser over the entire emission spectrum. As discussed in Section\,\ref{s:ICSHG}, the nonlinear output coupling is close to the optimum level over the entire emission wavelength range. The gain line center, where the output power is maximal, is found at around 671.1\,nm, resulting in 2.2\,W of fundamental and 2.1\,W of second-harmonic output. At the lithium-$D$ line wavelength the output power amounts to $\approx$1.8\,W. Being close to the gain line center, this result compares favorably to the value obtained in\,\cite{Camargo2010}, where the authors used an etalon for tuning which potentially produces significantly higher tilt loss\,\cite{Eismann2012a,Leeb1975}. Comparing the emitted power of the doubled and non-doubled lasers across the emission spectra, we typically obtain more than 80\% of the power at 671\,nm, demonstrating a very efficient frequency doubling process. For the absolute maximum values, we obtain $P_{2 \omega}/P_\omega = 88\%$. It is even possible to obtain emission further away from resonance, where the non-ICSHG lasers cease to oscillate. Our first-generation source presented in\,\cite{Eismann2012} displays an output spectrum that is significantly more plateau-like (gold triangles in Fig.\,\ref{f:tuningcurve}), which we attribute to the presence of a second, more selective etalon in the cavity. At the expense of a higher insertion loss baseline, ripples in the emission spectrum have significantly less influence on the tuning behavior of this setup. The identical gaps in all of the emission spectra presented here (A,B,C in Fig.\,\ref{f:tuningcurve}) can be explained by absorption from water vapor, as we compare the laser spectra to a water vapor absorption spectrum obtained from the HITRAN database\,\cite{Rothman2009}. The water absorption peaks coincide with the features A,B and C. 

\smallskip

Intracavity-frequency-doubled lasers tend to mode-locked operation when the doubling crystal is mismatched from the optimum phase matching condition\,\cite{DeSalvo1992,Holmgren2005}. This effect, resulting from an intensity-dependent phase shift of the fundamental beam passing the non-matched crystal, is commonly termed nonlinear-Kerr-lens- or $\chi^{(2)}:\chi^{(2)}$ mode locking. Although it has been observed in the 1064-nm-Nd:YVO$_4$-ppKTP system before\,\cite{Schafer2011}, we report, to the best of our knowledge, the first observation at 1342\,nm. 
We observe the effect when detuning from the optimum phase-matching temperature to below $T_{\rm low} = $29$^\circ$C and above $T_{\rm high} = $45$^\circ$C. A scanning Fabry-Perot spectrum analyzer shows no discernible single- or few-mode background once the threshold to mode-locked operation is crossed. We use a fast detector and a spectrum analyzer to measure the beat frequency between neighboring modes. This simple method to determine the laser cavity free spectral range yields a value of 345(1)\,MHz.

It is remarkable that the mode locking arises when tuning the phase-matching temperature to colder-than-optimal values. 
In analogy to standard Kerr-lens mode locking, the $\chi^{(2)}:\chi^{(2)}$ process requires an additional intracavity element which modifies the round-trip gain or loss as a function of the power-dependent cavity mode size. 
This can be realized by the gain medium, cf. Fig.\,\ref{f:laser}. As we presented in Section\,\ref{s:IRSetup}, the pump-to-cavity mode overlap has been carefully optimized in the cw regime. However, an ABCD-matrix-based cavity mode calculation reveals that the beam waist in the Nd:YVO$_4$  only changes monotonously when crossing over from negative to positive focal powers in the ppKTP. Thus, the induced change in gain or loss should be less favorable whenever the lensing departs from the optimized value. We note that in our simple analysis, we do not take into account different Kerr-like and thermal lenses that may occur in the intra-cavity elements other than the laser and nonlinear crystals. Using the temperature-dependent Sellmeier equations of \cite{Fradkin1999,Emanueli2003}, we get dephasing parameters of $\beta(T_{\rm low}) = 8.8(1)$ and $\beta(T_{\rm high}) = -18.5(1)$, resulting in a single-pass efficiency $\eta(T)$ reduced to less than 5\% of the maximum value.

\begin{figure}[htbp]
\centering\includegraphics[width=\linewidth]{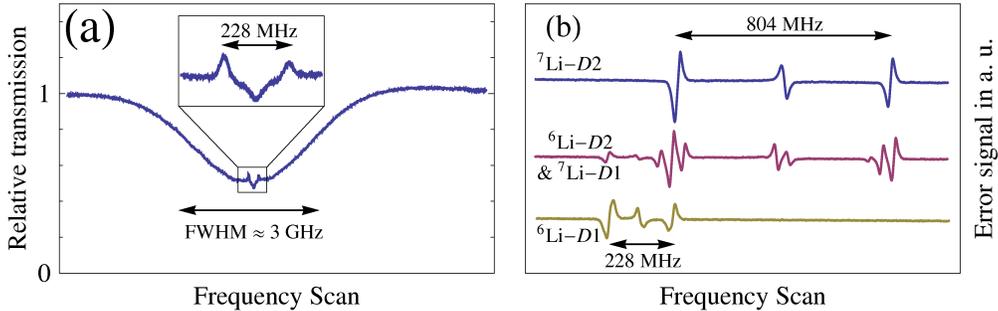}
\caption{(Color online) Doppler-free saturated absorption spectroscopy of the lithium D-lines. (a)~Sample scan over the entire $^6$Li $D1$ Doppler-broadened absorption peak. The inset shows the sub-Doppler features. Similar spectra are obtained for all lithium $D$~lines. (b)~Error signals of all lithium $D$-line transitions, generated by phase modulation spectroscopy. These signals provide an excellent reference for frequency-locking of the laser.}
\label{f:spectro}
\end{figure}

The laser can be tuned continuously using a piezoelectric transducer (PZT), upon which mirror M$_2$ is glued upon. This allows mode-hope free scans of the 671-nm output frequency over more than 6\,GHz. For the resonated fundamental light this amounts to about ten times the laser cavity free spectral range, a typical behavior of ICSHG lasers with self-suppressed mode hopping\,\cite{Martin1997,Helmfrid1994}.
We perform Doppler-free saturated absorption spectroscopy on a lithium vapor cell described previously in\,\cite{Eismann2012}, see Fig.\,\ref{f:spectro}. In Fig.\,\ref{f:spectro}(a) we show a sample scan over the full Doppler-broadened lithium-6~$D1$ line. 
The ground-state hyperfine structure is clearly resolved. We phase-modulate the probe beam at 20\,MHz using an electro-optic modulator, and then use a commercial lock-in amplifier to generate a dispersive error signal as presented in Fig.\,\ref{f:spectro}(b). We can access all $D$-line transitions of both naturally abundant lithium isotopes, and the full ground state hyperfine structure is resolved.

The laser linewidth was estimated by slowly scanning the emission frequency over the \mbox{$D1$-line} of \mbox{lithium-6.} From the observed peak-to-peak noise in the linear part of the error signal we obtain an upper limit of 1\,MHz for the laser linewidth. 
For frequency-locking the laser, we feed the error signal in a home-made proportional-integral controller. The regulator signal is split internally in a low-frequency part, sent to the slow PZT (M$_2$) used for scanning the laser, and a high-frequency part, sent directly to a fast PZT glued between M$_4$ and its mount. The system can easily stay locked for a full day, even in presence of significant acoustic noise coupled to the optical table. We note that due to the self suppression of mode hopping, it can be time consuming to optimize the laser output parameters after switch-off.

\section{Conclusion}

In conclusion, we presented the design and the performance of an all-solid-state, intracavity-frequency-doubled single-mode laser source. 
The ring laser emits up to 2.5\,W of unidirectional multifrequency radiation at the fundamental wavelength, and we determine the parameters necessary for efficient ICSHG. Starting from this well-characterized infrared system, the optimized ICSHG yields up to 2.1\,W of single-mode, frequency-stabilized output at 671\,nm. We discussed a simple theory describing the output power of an intra-cavity frequency-doubled laser. Within the region of efficient nonlinear conversion, the theory describes our experimental findings well. We furthermore presented a measurement of the emission spectrum of both the fundamental and the ICSHG source, displaying around 1.8\,W of output power at the lithium $D$-lines. By comparing both emission spectra, we maximally obtain 88\% of the fundamental power at 671\,nm, and typically more than 80\%. This demonstrates a very efficient ICSHG process across the full emission spectrum. Furthermore, we discussed the fine-tunability of the source by presenting Doppler-free saturated absorption spectra of all lithium $D$-line transitions. The 671-nm output can be mode-hop-free scanned over more than 6\,GHz, which corresponds to more than ten times the laser cavity free spectral range. As compared to the most powerful conventional design\,\cite{Eismann2012}, our laser is a significant improvement by more than a factor of three in terms of output power. Finally, we observe nonlinear-Kerr-lens mode-locked operation when detuning the ppKTP temperature sufficiently far from the phase matching condition.

\smallskip
The output power of our source is largely sufficient for creating all the laser beams needed for an ultracold atoms experiment. However, for other applications such as atom interferometry, even more power might be helpful and we will briefly discuss how this can possibly be established.
An obvious way is to use a different cavity geometry allowing for two-sided pumping of the active medium. Convex pump couplers can compensate the thermal lens close to the laser crystal, which has proved very efficient in high-power resonators, see for instance\,\cite{Lenhardt2009}. A large part of the output power limitations stems from the additional intracavity elements, such as the Faraday rotator. Thus, injection locking of an amplifier consisting of an empty high-power resonator such as presented in~\cite{Lenhardt2009} could deliver tens of watts of single-frequency fundamental light. At these elevated power levels, efficient single-pass frequency doubling of 1342-nm radiation has been demonstrated\,\cite{Lenhardt2010}, the implementation of which considerably relaxes the complexity of multi-cavity designs.

\section{Acknowledgements}

We acknowledge the group of Jacques~Vigu\'e for dicussions and exchange of material, and Carlota~Canalias for providing the ppKTP. The authors wish to thank Franz~Sievers for assembly of the lock electronics, and Colin~Parker, Peter~Scherpelz and Shih-Kuang~Tung for helpful comments on the manuscript.


\begin{thebibliography}{}
\newcommand{\enquote}[1]{``#1''}

\end{thebibliography}


\begin{thebibliography}{10}
\newcommand{\enquote}[1]{``#1''}

\bibitem{Giorgini2008}
S.~Giorgini, L.~P. Pitaevskii, and S.~Stringari, \enquote{Theory of ultracold
  atomic fermi gases,} Rev. Mod. Phys. \textbf{80}, 1215--1274 (2008).

\bibitem{Chin2010}
C.~Chin, R.~Grimm, P.~Julienne, and E.~Tiesinga, \enquote{Feshbach resonances
  in ultracold gases,} Rev. Mod. Phys. \textbf{82}, 1225--1286 (2010).
  
  \bibitem{Hou2011}
F.~Hou, L.~Yu, X.~Jia, Y.~Zheng, C.~Xie, and K.~Peng, \enquote{Experimental
  generation of optical non-classical states of light with 1.34\,$\upmu$m
  wavelength,} The European Physical Journal D 
  \textbf{62}, 433--437 (2011). 

\bibitem{Camargo2010}
F.~Camargo, T.~Zanon-Willette, T.~Badr, N.~Wetter, and J.~Zondy,
  \enquote{Tunable single-frequency Nd:YVO$_4$ BiB$_3$O$_6$ ring laser at
  671\,nm,} IEEE Journal of Quantum Electronics \textbf{46}, 804--809 (2010).
 
\bibitem{Eismann2012}
U.~Eismann, F.~Gerbier, C.~Canalias, A.~Zukauskas, G.~Tr\'enec, J.~Vigu\'e,
  F.~Chevy, and C.~Salomon, \enquote{An all-solid-state laser source at 671 nm
  for cold-atom experiments with lithium,} Applied Physics B \textbf{106},
  25--36 (2012). 

\bibitem{Fernandes2012}
D.~Fernandes, F.~Sievers, N.~Kretzschmar, S.~Wu, C.~Salomon, and F.~Chevy,
  \enquote{Sub-doppler laser cooling of fermionic $^{40}$K atoms in
  three-dimensional gray optical molasses,} arXiv preprint arXiv:1210.1310
  (2012).
  

\bibitem{Miffre2006}
A.~Miffre, M.~Jacquey, M.~B\"uchner, G.~Tr\'enec, and J.~Vigu\'e, \enquote{Atom
  interferometry measurement of the electric polarizability of lithium,} The
  European Physical Journal D - Atomic, Molecular, Optical and Plasma Physics
  \textbf{38}, 353--365 (2006). 

\bibitem{Muller2008}
H.~M\"uller, S.-w. Chiow, Q.~Long, S.~Herrmann, and S.~Chu, \enquote{Atom
  interferometry with up to 24-photon-momentum-transfer beam splitters,} Phys.
  Rev. Lett. \textbf{100}, 180405 (2008).

\bibitem{Olivares2002}
I.~E. Olivares, A.~E. Duarte, E.~A. Saravia, and F.~J. Duarte, \enquote{Lithium
  isotope separation with tunable diode lasers,} Appl. Opt. \textbf{41},
  2973--2977 (2002).
  
\bibitem{Payne1994}
S.~A. Payne, L.~K. Smith, R.~J. Beach, B.~H.~T. Chai, J.~H. Tassano, L.~D.
  DeLoach, W.~L. Kway, R.~W. Solarz, and W.~F. Krupke, \enquote{Properties of {Cr:LiSrAIF$_6$} crystals for laser operation}, Appl. Opt. {\bf 33},
  5526  (1994).
  
\bibitem{McDonagh2006}
L.~McDonagh, R.~Wallenstein, R.~Knappe, and A.~Nebel, \enquote{High-efficiency
  60\,W TEM$_{00}$ Nd:YVO$_4$ oscillator pumped at 888\,nm,} Opt. Lett. \textbf{31},
  3297--3299 (2006).

\bibitem{Tr'enec2011}
G.~Tr\'{e}nec, W.~Volondat, O.~Cugat, and J.~Vigu\'{e}, \enquote{Permanent
  magnets for faraday rotators inspired by the design of the magic sphere,}
  Appl. Opt. \textbf{50}, 4788--4797 (2011).

\bibitem{Laporta1991}
P.~Laporta and M.~Brussard, \enquote{Design criteria for mode size optimization
  in diode-pumped solid-state lasers,} IEEE Journal of Quantum Electronics
  \textbf{27}, 2319--2326 (1991).

\bibitem{Chen1997}
Y.~F. {Chen}, T.~M. {Huang}, C.~F. {Kao}, C.~L. {Wang}, and S.~C. {Wang},
  \enquote{{Optimization in scaling fiber-coupled laser-diode end-pumped lasers
  to higher power: influence of thermal effect},} IEEE Journal of Quantum
  Electronics \textbf{33}, 1424--1429 (1997).

\bibitem{Rigrod1963}
W.~Rigrod, \enquote{Gain saturation and output power of optical masers,}
  Journal of Applied Physics \textbf{34}, 2602--2609 (1963).

\bibitem{Khazanov2004}
E.~Khazanov, N.~Andreev, A.~Mal'shakov, O.~Palashov, A.~Poteomkin, A.~Sergeev,
  A.~Shaykin, V.~Zelenogorsky, I.~Ivanov, R.~Amin \emph{et~al.},
  \enquote{Compensation of thermally induced modal distortions in Faraday
  isolators,} IEEE Journal of Quantum Electronics \textbf{40}, 1500--1510
  (2004).

\bibitem{Lenhardt2009}
F.~Lenhardt, M.~Nittmann, T.~Bauer, J.~Bartschke, and J.~L'huillier,
  \enquote{High-power 888-nm-pumped  Nd:YVO$_4$ 1342-nm oscillator operating in
  the TEM$_{00}$ mode,} Applied Physics B \textbf{96}, 803--807 (2009).

\bibitem{Smith1970}
R.~Smith, \enquote{Theory of intracavity optical second-harmonic generation,}
  IEEE Journal of Quantum Electronics \textbf{6}, 215--223 (1970).

\bibitem{Boyd1968}
G.~D. Boyd and D.~A. Kleinman, \enquote{Parametric interaction of focused
  gaussian light beams,} Journal of Applied Physics \textbf{39}, 3597--3639
  (1968).

\bibitem{Fradkin1999}
K.~Fradkin, A.~Arie, A.~Skliar, and G.~Rosenman, \enquote{Tunable midinfrared
  source by difference frequency generation in bulk periodically poled
  KTiOPO$_4$,} Applied Physics Letters \textbf{74}, 914--916 (1999).

\bibitem{Emanueli2003}
S.~Emanueli and A.~Arie, \enquote{Temperature-dependent dispersion equations
  for KTiOPO$_4$ and KTiOAsO$_4$,} Appl. Opt. \textbf{42}, 6661--6665 (2003).

\bibitem{Peltz2001}
M.~Peltz, U.~B\"ader, A.~Borsutzky, R.~Wallenstein, J.~Hellstr\"om,
  H.~Karlsson, V.~Pasiskevicius, and F.~Laurell, \enquote{Optical parametric
  oscillators for high pulse energy and high average power operation based on
  large aperture periodically poled KTP and RTA,} Applied Physics B
  \textbf{73}, 663--670 (2001). 

\bibitem{Eismann2012a}
U.~Eismann, \enquote{A novel all-solid-state laser source for lithium atoms and
  three-body recombination in the unitary bose gas,} Ph.D. thesis, Universit\'e
  Pierre et Marie Curie -- Paris VI (2012).

\bibitem{Martin1997}
K.~I. Martin, W.~A. Clarkson, and D.~C. Hanna, \enquote{Self-suppression of
  axial mode hopping by intracavity second-harmonic generation,} Opt. Lett.
  \textbf{22}, 375--377 (1997).

\bibitem{Helmfrid1994}
S.~Helmfrid and K.~Tatsuno, \enquote{Stable single-mode operation of
  intracavity-doubled diode-pumped Nd:YVO$_4$ lasers: theoretical study,} J. Opt.
  Soc. Am. B \textbf{11}, 436--445 (1994).

\bibitem{Leeb1975}
W.~Leeb, \enquote{Losses introduced by tilting intracavity etalons,} Applied
  Physics A: Materials Science \& Processing \textbf{6}, 267--272 (1975). 

\bibitem{Rothman2009}
L.~Rothman \textit{et al.},
\enquote{The HITRAN 2008 molecular
  spectroscopic database,} Journal of Quantitative Spectroscopy and Radiative
  Transfer \textbf{110}, 533 -- 572 (2009).

\bibitem{DeSalvo1992}
R.~DeSalvo, D.~J. Hagan, M.~Sheik-Bahae, G.~Stegeman, E.~W.~V. Stryland, and
  H.~Vanherzeele, \enquote{Self-focusing and self-defocusing by cascaded
  second-order effects in KTP,} Opt. Lett. \textbf{17}, 28--30 (1992).

\bibitem{Holmgren2005}
S.~Holmgren, V.~Pasiskevicius, and F.~Laurell, \enquote{Generation of 2.8-ps
  pulses by mode-locking a Nd:GdVO$_4$ laser with defocusing cascaded kerr lensing
  in periodically poled KTP,} Opt. Express \textbf{13}, 5270--5278 (2005).

\bibitem{Schafer2011}
C.~Sch\"{a}fer, C.~Fries, C.~Theobald, and J.~A. L'huillier,
  \enquote{Parametric kerr lens mode-locked, 888\,nm pumped Nd:YVO$_4$ laser,} Opt.
  Lett. \textbf{36}, 2674--2676 (2011).
  
\bibitem{Lenhardt2010}
F.~Lenhardt, A.~Nebel, R.~Knappe, M.~Nittmann, J.~Bartschke, and J.~A.
  L'huillier, \enquote{Efficient single-pass second harmonic generation of a
  continuous wave Nd:YVO$_4$- laser at 1342\,nm using MgO:ppLN,} in
  \enquote{Conference on Lasers and Electro-Optics,}  (Optical Society of
  America, 2010), p. CThEE5.

\end{thebibliography}
\end{document}